# The physical foundations for the geometric structure of relativistic theories of gravitation. From General Relativity to Extended Theories of Gravity through Ehlers-Pirani-Schild approach

S. Capozziello[1,2], M. De Laurentis[1,2], L. Fatibene[3,4], M. Francaviglia [3,4]
[1]*Dipartimento di Scienze Fisiche, Università di Napoli "Federico II",*
*Compl. Univ. di Monte S. Angelo, Edificio G, Via Cinthia, I-80126 - Napoli, Italy*
[2]*INFN Sez. di Napoli, Compl. Univ. di Monte S. Angelo, Edificio G, Via Cinthia, 80126 Naples, Italy*
[3]*Dipartimento di Matematica, Università di Torino, Turin, Italy*
[4]*INFN Sez. di Torino, Via Carlo Alberto 10, 10123 Turin, Italy*
(Dated: November 1, 2018)

We discuss in a critical way the physical foundations of geometric structure of relativistic theories of gravity by the so-called Ehlers-Pirani-Schild formalism. This approach provides a natural interpretation of the observables showing how relate them to General Relativity and to a large class of Extended Theories of Gravity. In particular we show that, in such a formalism, geodesic and causal structures of space-time can be safely disentangled allowing a correct analysis in view of observations and experiment. As specific case, we take into account the case of $f(R)$ gravity.

## I. INTRODUCTION

Einstein's General Relativity (GR) is a self-consistent theory that dynamically describes Space, Time and Matter under the same standard. The result is a deep and beautiful scheme that, starting from some first principles, is capable of explaining a huge number of gravitational phenomena, ranging from laboratory up to cosmological scales. Its predictions are well tested at Solar System scales and it gives rise to a comprehensive cosmological model that agrees with the Standard Model of particles, the recession of galaxies, the cosmic nucleosynthesis and so on. This scheme will be called the *standard GR* framework.

Despite these good results, the recent advent of the so-called Precision Cosmology entails that the self-consistent scheme of GR seems to disagree with an increasingly high number of observational data, as *e.g.* those coming from IA-type Supernovae, used as standard candles, large scale structure ranging from galaxies up to superclusters [1–3]. Next generation of gravitational experiments can also produce similar mismatch between theory and observations at Solar System level.

Furthermore, being not renormalizable, GR seems to fail to be quantized in any standard way (e.g. see [4]). In other words, it seems, from ultraviolet up to infrared scales, that GR is not and cannot be the final theory of gravitation even if it successfully addresses a wide range of phenomena. Of course there are some promising proposals for quantizing gravitational sector in a non-standard way (*e.g.* string theory, loop quantum gravity or dynamical triangulation [5–8]). Probably it is worth noticing that in most cases these frameworks for quantum gravity do in fact imply classical modified dynamics in the weak field limit (see [9–11]).

Many attempts have been therefore made both to recover the validity of GR at all scales, on one hand, and to produce theories that suitably generalize Einsteins standard GR, on the other hand. In order to interpret a large number of recent observational data inside the paradigm of GR, the introduction of Dark Matter (DM) and Dark Energy (DE) seemed to be necessary: the price of preserving the *simplicity* of the Hilbert Lagrangian has been, however, the introduction of rather odd-behaving physical entities which, up to now, have not been revealed by any experiment at fundamental scales. In other words, we are observing the large scale effects of missing matter (DM) and the accelerating behavior of the Hubble flow (DE) but no final evidence of new ingredients exists, if we want to deal with them as standard quantum particles or fields [12].

In particular, when the expansion rate of the Universe is enhanced (as compared to that one derived in the framework of GR), it is observed that thermal relics decouple with larger relic abundance. The Hubble rate change could have its imprint on the relic abundance of DM, such as WIMPs, axions, heavy neutrinos. This kind of studies are motivated by recent astrophysical results which involve cosmic ray electron and positrons [13–16], antiprotons [17], and $\gamma$-rays [18, 19]. Specifically the rising behavior of the positron fraction observed in PAMELA (Payload for Antimatter Matter Exploration and Light-nuclei Astrophysics) experiment [13] could be the "signature" of new physics that cannot be framed in the standard scheme of GR. Beside the astrophysical interpretation of this phenomenon [20], it is under investigation the possibility that the raise of cosmological scale factor could be due to dark matter annihilation, dominantly occurred into lepton [21, 22]. In this last case, it is required a large value of $\langle\sigma_{ann}v\rangle$. More specifically,

PAMELA and ATIC (Advanced Thin Ionization Calorimeter) data require a cross section of the order or larger than $\langle \sigma_{ann} v \rangle \sim 10^{26}$ cm$^3$ sec$^{-1}$: in this case, thermal relics could be interpreted either as the observed DM density [23] or as the signature for new physics.

In any event, at least from an observational point of view, considering GR+cosmological constant+DM gives a good snapshot of the currently observed Universe. However, the problem with this framework is that dynamics of previous epochs cannot be reconstructed and addressed in a self-consistent and satisfactory way starting from the present status of observations. Furthermore, it seems that the type of DM to be considered strictly depends on the size of self-gravitating structures (*e.g.* the dynamical behavior of DM in small galaxies, in giant galaxies and in galaxy clusters is completely different). Thus, besides the issue to find out DM and DE at fundamental level, it seems hard to find out a general dynamics involving such components working well at all cosmic epochs and at any astrophysical scales.

With these considerations in mind, one can wonder if extending gravity sector to non-standard dynamics could be a more economic and useful approach which does not involve too much exotic ingredients but retains all the good results achieved by standard GR (see *e.g.* [24–30]).

In this paper, we want to address some of the recent issues concerning the geometrical structure of "physically reasonable" gravitational theories, starting from the fundamental work of Elehers-Pirani-Schild (EPS) [31, 32] about the geometric and physical foundations of relativistic theories of gravitation and revisiting them, *á la Palatini*, in view of applications to the new challenges discussed above [35, 36].

The outline of the paper is as follows. In Section II we introduce the EPS framework. Section III is devoted to the EPS formalism in GR while Section IV is a critical discussion of such an approach. In Section V, we discuss EPS from the point of view of Extended Theories of Gravity (ETG). In particular, the straightforward extension of GR, $f(R)$-gravity, is taken into account. Section VI is devoted to discussion and conclusions. A new paradigm for gravitational theories is proposed assuming the EPS paradigm.

## II. THE EHLERS-PIRANI-SCHILD APPROACH

Let us first summarize the EPS analysis of the mathematical structures that lie at the basis of all "reasonable relativistic theories of gravity" [31].

In early 70s, EPS started from a set of well motivated physical properties of light rays and matter worldlines in a relativistic framework to derive the geometrical structure of space-time from potentially observable objects. This is particularly suitable to discuss which geometric structures are observable and which are conventional.

In this way, EPS formalism provides stronger physical motivations and understanding not only of space-time geometry as such, but also in comparison with more general geometries (as candidates for mathematically modeling physical space-time). In fact, EPS highlighted the potential role of space-time models based on Weyl geometries, which contain standard Lorentzian models as a special case. Let us also stress that observation protocols in GR strongly rely on the assumptions on geometry of space-time and (at least) on the physical properties of electromagnetic field. Moreover, the wider EPS-Weyl framework potentially allows to think about experiments which can potentially test the assumptions of GR directly, not only by the indirect consequences on motions of matter and light. Supplying this new axiomatic characterization of the otherwise mathematically familiar space-time geometry structure, EPS formalism also brings relevant new insight even from a strictly mathematical (geometrical) standpoint.

Einstein's standard GR uses advanced mathematical ideas. Objects like 4-dimensional curved space-time are not easy to grasp. Even if one masters the math behind it, the essential physical meaning and content is not obvious. EPS is one of a series of attempts to clarify the physics behind the math. Unfortunately and unavoidably, getting there requires even more abstract math. At first sight, this seems self-defeating; however, some of these mathematical ideas are chosen so as to be closer to the 'operational' physical interpretation, representing the most elementary physical observables, measurements and constructions.

In the upshot, EPS ends up with something like the familiar Lorentzian metric ($L4$), rather then of accepting it as starting point [37]. To be precise, EPS shows that one can obtain a class of Lorentzian metrics from lightrays, *i.e.* by analyzing the light behavior. The class obtained is a *conformal structure*, namely it contains all the metrics differing by a pointwise conformal factor, *i.e.* in the form $g'_{\mu\nu} = \varphi(x) \cdot g_{\mu\nu}$ The idea is to rebuild $L4$ from scratch, using only bricks with intuitively clear physical observational meaning to the extent possible, and at the cost of some extra math. Since one gets a conformal class, that means that one cannot observe a representative of conformal gauge, *i.e.* a specific metric, but only the conformal class can be observed. In other words, a specific representative of conformal gauge can be chosen only on a conventional basis and physical observables should not depend on the representative but on the conformal structure only. For example, the lightcones are conformally invariant and are hence eligible for being observable. Let us stress that conformal structure is build essentially out of light rays and a further structure (see below) is needed to describe matter free fall.

As is typical in axiomatic reconstructions like EPS, one exploits the benefit of hindsight, as the intended result (in this case: $L4$ space-time of GR) is already known. So this in no way detracts from Einstein's original feat, on the contrary. The scope of EPS is limited to the kinematics of space-time itself; the problem of any possible axiomatic derivation or reconstruction of Einstein field equations (*i.e.* dynamics) governing matter and gravity within such a space-time model, is left open. Quoting Ehlers [31–33]:

*The approach shows how quantitative measures of time, angle and distance, and a procedure of parallel displacement... can be obtained constructively from 'geometry free' assumptions about light-rays and freely falling particles; pseudo-Riemannian (or Weylian) geometry is recognized even more clearly than before as the appropriate language for a generalized kinematics which allows for the unavoidable and ever-present 'distortions "called gravitational fields".*

With these considerations in mind, let us outline the main points of EPS conceptual construction.

### A. Outline of the Ehelers-Pirani-Schild formalism

The construction of EPS space-time proceeds in steps as sketched below, each one enriching the axiomatic content of the underlying set of events. Roughly speaking, the underlying idea is the following. From differential geometry, one knows that the geodesics determine "their affine connection" (assuming torsion to be zero, for instance). Let us stress the important point that this connection is not imposed to be metric.

Now, in contrast to the metric (or connection) itself, these geodesics do possess an immediate physical interpretation (as light ray worldlines for null geodesics or particle world lines for timelike ones). So in very general terms, one tries to reconstruct the sought-after metric from known families of worldlines that fulfill certain qualitative criteria (postulates), which are themselves physically meaningful and plausible.

- **Particles and light rays in event space**.

  EPS adopts a set $\mathcal{M}$ of events (to become the space-time manifold) as its backdrop. On this, the set of worldlines of particles $p$ and the set of light rays $l$ are assumed given. Both families are a collection of trajectories on $\mathcal{M}$.

- **Smooth radar coordinates for events**

  As subsets of the space of events, particle and light worldlines are assumed to be smooth one dimensional submanifolds in $\mathcal{M}$. One can use light rays and particles to define local coordinates on space-time mimicking a simple GPS system. These local coordinates are preferred in view of their relation with standard observation protocols and deserve a special role in a physical model of space-time. A *permissible local coordinate system* represents time as measured by a (possibly irregular) local clock. Light ray messages between particles $p$ and $q$ smoothly relate their private time parameters, the timing of echoes received back by $p$ also relate smoothly to that of the message flashes it sent out to $q$ to begin with. Using 'radar soundings' in this way, pairs of 'observer' particles set out to map surrounding events by assigning 2 time values each, or a total of 4 coordinates each. Of course, once these radar coordinates are defined then one is free to use any compatible coordinate system.

- **Light propagation ensures local validity of pointwise causality**

  At each point of space-time (event), the propagation of light determines an infinitesimal null cone, amounting to a conformal structure $\mathcal{C}$ of Lorentzian signature. In particular, one can define the conformal class of Lorenztian metrics and show *a posteriori* that light cones split vectors into $\mathcal{C}$-time-like, space-like and null vectors. Usually, timelike vectors are defined in view of their norm with respect to the space-time metric (or conformal structure). Here they are *defined* (topologically) by the distribution of lightcones and then *proven* to have the correct standard norm to characterizing them. This assertion is stated operationally, demanding that one may (topologically) distinguish between $\mathcal{C}$-time-like, space-like and null vectors, directions and curves at an event. In the same way, one can show that null curves lying on a null hypersurface (namely the light cones) are singled out as null geodesics of (any) representative of the conformal structure.

- **Free falling particles encode influence of gravity on particle motion**

  Now that timelike is defined, the free-falling particles form a preferred family of wordlines among the timelike curves. Imposing a generalized law of inertia (or resorting to general covariance requirements) provides a *projective structure* [34]. A projective structure $\mathcal{P}$ is a class of connections which share the same worldlines as geodesics. Each representative only parametrize differently the same geodesics trajectories. Let us stress that nothing indicates, at this level, that such connections need to be metric connections. Then particles are identified with free-fall worldlines as ($\mathcal{C}$-time-like) $\mathcal{P}$-geodesics.

- **Light and particle motion agree**

  In line with physical experience, one assumes that particles can be made to chase a photon arbitrarily closely, meaning their paths "fill the light cone" so that each $\mathcal{C}$, null geodesic is also a $\mathcal{P}$ geodesic. This compatibility condition links the conformal structure with the connection and can be shown to be equivalent to require that $\mathcal{C}$-null geodesics (namely light rays) are a subset of $\mathcal{P}$-geodesics. This is in turn equivalent to being able to split the set of $\mathcal{P}$-geodesics into timelike, spacelike and null trajectories. If one geodesic has a character at a point it maintains the same character all along (as it happens in standard GR in view of metricity). Still in view of this compatibility condition the connection used to describe free fall is more general than the one induced by the metric structure alone.

- **Free fall connection**

  Then one can define two compatible conformal and projective structures on space-time. The choice of representatives is a conventional gauge fixings. The conventional nature of metrics and connections is important to be noticed in view of which quantities are to be considered physically sound. In particular, one can choose canonically a standard representative of projective structure imposing

  $$\nabla_\mu^{(\Gamma)} g_{\alpha\beta} = 2V_\mu g_{\alpha\beta} \tag{1}$$

  for some covector $V$. Then there is a canonical connection $\Gamma^\alpha_{\beta\mu}$ which, of course, depends on extra degrees of freedom depending on $A$. The triple $(\mathcal{M}, \mathcal{C}, \Gamma)$ is called a *Weyl-geometry*. It is called *metric* if there exists a representative $g \in \mathcal{C}$ such that $\Gamma = \{g\}$ coincides with the Christoffel symbols of the metric $g$. In this case, the metric $g$ describes light rays and particles free fall, as it is assumed in standard GR. However, in general one needs two different (still compatible) structures to describe light rays and matter free fall. Let us stress once again that there is no reason at this stage to assume that the Weyl-geometry obtained on space-time is metric. A Weyl space possesses a unique affine structure $\mathcal{A}$: $\mathcal{A}$ geodesics are $\mathcal{P}$ and $\mathcal{A}$ parallel displacement preserves $\mathcal{C}$ nullity. In a Weyl space, one may construct a "proper time" arc length (up to linear transformation) along non-null curves by purely geometrical means (*i.e.* using light rays reflected from particles only, so without any need for atomic clocks). In technical terms, one employs affine parallel displacement, and congruence in the tangent space, as defined by $\mathcal{C}$. This 'geodesic' clock is known as the Marke-Wheeler clocks [38].

### B. Hypotheses and assumptions

In summary EPS analysis is based on a number of assumptions:

It physically distinguishes the Principle of Equivalence from the Principle of Causality and investigates the need of measuring and describing space-time structure through light rays and particles. The need of measuring in space-time and using light rays requires that space-time carries (a class of Lorentzian) metrics, while the Principle of Equivalence and interaction with matter ("Free Fal" under gravitational pull) requires that space-time carries also a (linear or affine) connection.

The connection, an object that can be reduced to be zero at each single point, is the potential of the gravitational field. The metric determines causality and photon propagation. According to EPS analysis, in order for a gravitational theory being physically reasonable, compatibility conditions should exist between the metric and the connection. The connection defines a family of autoparallel lines (also called improperly geodesics). They establish the free fall of pointlike (in principle massive) "test particles". The metric defines light cones and a family of geodesics. Null geodesics of the metric are paths of light rays (photons). The family of autoparallel lines of the connection determine an equivalence class of "Projectively Equivalent Connection". Along them, free fall is the same, only proper time changes. The light cones of the metric define an equivalence class of conformally equivalent metrics. Along them units and measuring devices change point by point, but light rays and photon trajectories are the same. The required compatibility condition amounts to pretend that the two families of autoparallel lines of the (projective equivalence class of) connections and the family of null geodesics of the conformal equivalence class of metrics are in a precise relation: each null geodesic of the metric has to be one of the autoparallel lines of the connection At this point of their fresh analysis, all foundational Axioms have been satisfied. In order to recover GR as the Unique Relativistic Theory of Gravitation Elhers, Pirani and Schild made some further axiomatic hypotheses:

- **Speed of time does not depend on path**

  A final physical assumption (expressed mathematically as an axiom) ensures the existence of a Lorentzian metric, which determines both light cones and free fall.

"Equally spaced clock ticks" along one particle world line are transported to a nearby particle by Einstein simultaneity. Imposing that this must generate (approximately) equidistant ticks also for the second particle and applying the equation of geodesic deviation for the curvature tensor given by $\mathcal{A}$ implies (through the vanishing of the Weyl 'track curvature') the existence of a single Lorentzian metric compatible to both $\mathcal{C}$ with $\mathcal{A}$.

This finally 'reduces' Weyl space to $L4$. Requiring in this way that 'time runs equally fast along all paths' amounts to denying the existence of a 'second clock effect'. Indeed, in (Lorentzian) GR, only the 'time interval' between 2 events is path dependent (*i.e.* the 'first clock effect'); not the 'speed' of time.

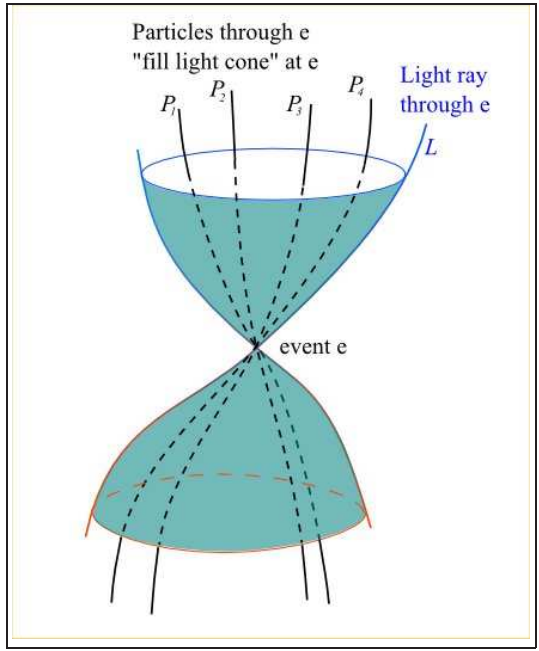

FIG. 1: Free particle (projective) geodesics "fill" light ray (conformal) geodesic light cone at any event e, thus forming a Weyl manifold.

- *"Metricity Axiom"*: a single metric is chosen in the conformal class and the connection is chosen while be the Levi-Civita connection of this metric.

With this hypothesis, the compatibility conditions are met. Notice, however, that this just amounts to say that the gravitational theory is of "purely metric" nature. To recover GR as the Unique Relativistic Theory of Gravitation one has in fact to make a further assumption.

### III. THE ELEHERS-PIRANI-SCHILD APPROACH FOR GENERAL RELATIVITY

In order to recover GR as the Unique Relativistic Theory of Gravitation, one has in fact to make the following further axiomatic hypotheses:

**"Lagrangian Axiom"**: the Lagrangian that governs gravitational field equations (in absence of matter) is the scalar curvature $R$.

The **"Metricity Axiom"** has in fact no real physical grounds. According to EPS (and to physical needs) a metric has to exist to define rods and clocks, but there is no need to pretend from the very beginning that it defines also the gravitational potential, *i.e.* the connection. Assuming that the Metricity Axiom holds is just a "matter of taste" and, in a sense, it corresponds to have a great mathematical simplification. From the viewpoint of Lagrangian Mechanics, it is a purely kinematical restriction imposed *a priori* on Dynamics. Physically speaking, it is much better not to impose *a priori* purely kinematical restriction on Dynamics. Physics requires that possible restrictions should be obtained from dynamics rather than imposed *a priori* as a constrains. This point was perfectly clear to Albert Einstein when, in 1923, he tried to establish a more general setting for Gravity (and Electromagnetism) by assuming *a priori* that both a metric and a connection must be chosen, from the beginning, as dynamical variables. So-called "Palatini formalism" was born.

On the other hand, also the "Lagrangian Axiom" had in fact no real physical grounds. Once it is clear which are the variables that have to enter dynamics, the choice of a Lagrangian for them is again a "matter of taste" or it should be at least determined on the basis of phenomenology, in order to fit observational data. When Hilbert, in 1916, in the purely metric framework (the only one that was available before 1919 and Levi-Civita's work on Linear Connections) assumed the Lagrangian to be the scalar curvature of the metric this was, in a sense, an obliged choice [39]. Dictated by "simplicity". The choice of the Hilbert-Einstein Lagrangian $R(g)$, made in 1916, was not only the "simplest one" [40–42]. It satisfied the will to obtain second-order field equations suitably generalizing Newtons law, fitting all astronomical predictions, satisfying conservation of matter and being compatible with Maxwell theory. The well known result is the Einstein field equations

$$G_{\alpha\beta} = R_{\alpha\beta} - \frac{1}{2} R \, g_{\alpha\beta} = k T_{\alpha\beta} \,, \tag{2}$$

where $G_{\alpha\beta}$ is the Einstein tensor, a combination of curvature invariants derived from Bianchi's identities, $T_{\alpha\beta}$ is the stress-energy momentum tensor and $\kappa = 16\pi G$ is the gravitational coupling constant.

In the new framework introduced by assuming both metric and connection among the variables, Einstein decided to take into account again, in 1923, the Lagrangian to be the scalar curvature (of metric and connection), again for the sake of simplicity. At that time there were very few observations fine enough to be used as tests and all of them agree with purely metric predictions. Thus the first test for an extended theory was to reproduce standard GR in purely metric formalism.

In this new framework and with the Linear Lagrangian $R(g, \Gamma)$ Einstein proved that no really new Physics comes on stage (in vacuum or with minimal matter couplings). Field equations impose, *a posteriori*, that the connection is nothing else but the Levi-Civita connection of the metric, so that GR is eventually recovered. Einstein did not investigate, however, what happens when matter is non-minimally coupled to gravity. In this case just a few slight changes are necessary if matter couples with metric $g$ but great difficulties arise if matter couples with the connection $\Gamma$ (as it should be if one has to take into account EPS formalism seriously). Around the sixties a number of mathematical papers were written about possible generalizations of Einstein's standard GR by reverting to non-linear Lagrangians, more complicated than $R(g)$. These Extended Theories remained just as a mathematical game for long time. Renewed interest towards non-linear Lagrangians was lately determined by new phenomenology, such as: Inflation, Acceleration in the Expansion, DM, Quantum Gravity, Low Energy Limit of String Models [29, 30].

## IV. REVISITING THE ELEHERS-PIRANI-SCHILD FORMALISM

The EPS analysis (concerning the mathematical and physical foundations of relativistic theories of gravitation and the compatibility between (conformal classes of) metrics and (projective classes of) connections) is worth of being revisited in view of ETG. EPS, in fact, provides an interpretation of gravitational theory which strongly constrains both kinematics and dynamics.

EPS have shown that the family of gravitational theories that satisfy all of their Axioms (with the exception of the "Metricity Axiom" and the "Lagrangian Axiom") includes most (but not all) of the currently investigated frameworks for (relativistic) gravitation.

Moreover, EPS formalism clearly suggests that, first of all, the correct and most general framework for dealing with gravity is the Palatini formalism, since it is based on the physical and mathematical distinction between the Principle of Equivalence and the Principle of Causality that for obvious reasons are mathematically and physically distinct. They imply the necessity of introducing *a priori* distinct and separate structures to full fill them, even if compatibility is required *a posteriori* on the mathematical and physical structures they induces on space-time.

EPS also imposes constraints on dynamics and matter couplings. In fact, it prescribes that matter free fall is governed by the connection and accordingly matter should couple with the connection. Moreover, the compatibility imposes that couplings are carefully chosen so that the compatibility of metric and connections can be dynamically enforced and maintained.

Within this formalism, the most general class of theories that can be considered according to the physical requirements pointed out by EPS analysis, is the family of so-called *Further Extended Theories of Gravity* that has been explicitly introduced in [43–47]. This class includes all gravitational theories in which the gravitational Lagrangian depend on $g$ and the (Ricci) curvature of the connection, the matter Lagrangian allows in principle interactions of matter with both $g$ and $\Gamma$ and, *a posteriori*, field equations imply EPS compatibility. Of course one is free to work in more general frameworks for gravitation, but in such a case one has to remind that at least one EPS requirements will fail or need to be modified.

Our choice wil be more restrictive and will be therefore based on three assumptions:

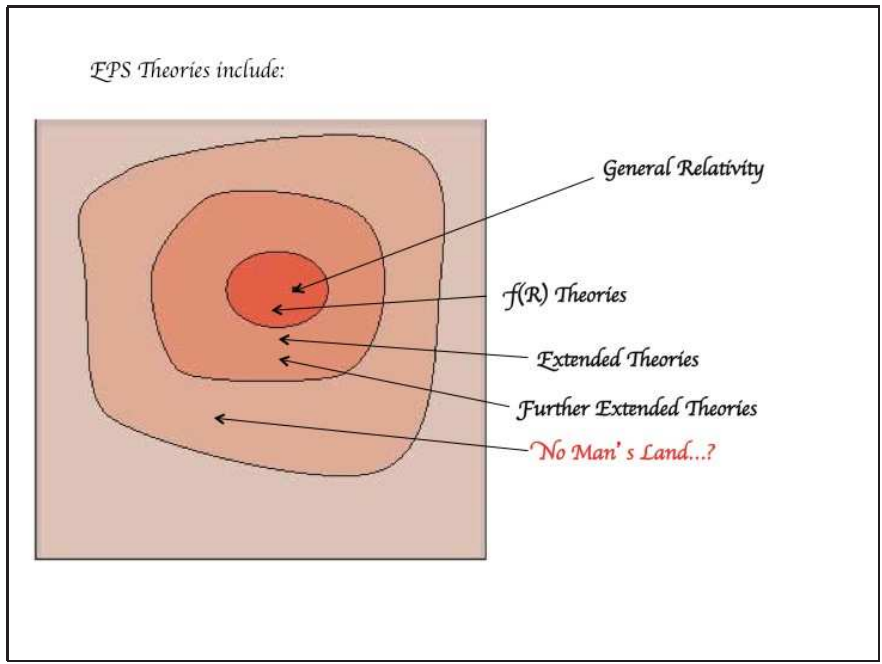

FIG. 2: the EPS theories

1. assume Palatini - EPS framework and accept the view that, in Palatini formalism, the gravitational field is encoded by a dynamical connection (*i.e.* free fall) while the dynamical metric has more to do with measures, rods, clocks and causality. We also accept that dynamics, and in particular the interaction with matter, will determine *a posteriori* the relations between the metric and the connection, so to satisfy EPS compatibility requirements.

2. assume that the Lagrangian is a possibly non-linear function of curvature of $g$ and $\Gamma$;

3. assume that the Lagrangian is "simple" - the simplest choice is of course $f(R(g,\Gamma))$ but it is not the only simple case.

This class of theories has extra features with respect to generic further Extended Theories of Gravity. In fact one can show that the connection is *a posteriori* forced to be metric in view of field equations. It is however, "metric" of a different metric $\hat{g}$, conformally equivalent to the metric $g$ originally chosen to represent rods and clocks. In this formalism for $f(R)$ theories then one has a natural mismatch between the canonical metric structure of space-time $\hat{g}$ which determines both light cones and free fall, and the metric $g$ used for rods and clocks. For this reason a deep review of observational protocols is needed in order to compare theory with observational tests (see for example the space mission proposal STE-QUEST, Space-Time Explorer and Quantum Equivalence Principle Space Test, [48]).

## V. THE EHELERS-PIRANI-SCHILD APPROACH FOR EXTENDED GRAVITY: THE CASE OF $f(R)$-GRAVITY

With this choice one can show that if no matter is present, purely gravitational equations entail that the connection $\Gamma$ entering dynamics is still the Levi-Civita connection of the metric $g$ while a (quantized) Cosmological Constant enters the game and somehow determines the asymptotic freedom for gravity.

One should remark that if matter is present and couples only to the metric $g$ things change if and only if the trace $T$ of the stress-energy tensor is different from zero. In particular, thence, nothing changes when electromagnetic couplings are present (being electromagnetic tensor traceless), so that the light cone structure and photons are not affected when switching to Palatini framework.

It is however known that - both in purely metric formalism (higher order gravity) and in the Palatini approach (first order gravity) - coupling with matter generates relativistic effects that are not present in vacuum. This is particularly evident when one relies on non-linear Lagrangians of $f(R)$-type. In $f(R)$ gravity (in the Palatini approach) in presence of matter coupled only with $g$ field equations still imply that the connection is metric, but now it is the Levi-Civita Connection of a new metric $\hat{g}$, conformally related with the metric $g$ given in the Lagrangian.

Being this the core point of our discussion, we want to derive in details the field equations of $f(R)$ gravity in Palatini formalism and then perform the EPS analysis in this framework.

Let us first consider on $\mathcal{M}$ a metric field $g$, a torsionless connection $\Gamma$ and a generic tensor density $A$ of rank 1 and weight $-1$. The covariant derivative of $A_\mu$ is then defined as

$$\overset{\Gamma}{\nabla}_\mu A_\nu = d_\mu A_\nu - \Gamma^\lambda_{\nu\mu} A_\lambda + \Gamma^\lambda_{\lambda\mu} A_\nu \,, \tag{3}$$

Accordingly, we have

$$\overset{\Gamma}{\nabla}_{(\mu} A_{\nu)} = d_{(\mu} A_{\nu)} - \left(\Gamma^\epsilon_{\nu\mu} - \delta^\epsilon_{(\nu}\Gamma^\lambda_{\mu)\lambda}\right) A_\epsilon = d_{(\mu} A_{\nu)} - u^\epsilon_{\mu\nu} A_\epsilon \,, \tag{4}$$

where we set $u^\epsilon_{\mu\nu} := \Gamma^\epsilon_{\mu\nu} - \delta^\epsilon_{(\mu}\Gamma^\lambda_{\nu)\lambda}$.

Let us consider the following Lagrangian (density)

$$\mathcal{L} = \frac{1}{\kappa}\sqrt{g}f(R) + gg^{\mu\nu}\overset{\Gamma}{\nabla}_\mu A_\nu \,, \tag{5}$$

where $g = |\det(g_{\mu\nu})|$, $R = g^{\mu\nu} R_{\mu\nu}(\Gamma)$ is the scalar curvature of $(g,\Gamma)$, and $f(R)$ is a generic (analytic) function. By variation of this Lagrangian and usual covariant integration by parts one obtains

$$\begin{aligned}
\delta\mathcal{L} &= \frac{\sqrt{g}}{\kappa}\left(f'(R)R_{(\alpha\beta)} - \frac{1}{2}f(R)g_{\alpha\beta} - \kappa T_{\alpha\beta}\right)\delta g^{\alpha\beta} + \\
&\quad - gg^{\alpha\beta}A_\lambda \delta u^\lambda_{\alpha\beta} + \frac{\sqrt{g}}{\kappa}g^{\alpha\beta}f'(R)\overset{\Gamma}{\nabla}_\lambda \delta u^\lambda_{\alpha\beta} + gg^{\mu\nu}\overset{\Gamma}{\nabla}_\mu \delta A_\nu = \\
&= \frac{\sqrt{g}}{\kappa}\left(f'(R)R_{(\alpha\beta)} - \frac{1}{2}f(R)g_{\alpha\beta} - \kappa T_{\alpha\beta}\right)\delta g^{\alpha\beta} - \frac{1}{\kappa}\left(\overset{\Gamma}{\nabla}_\lambda\left(\sqrt{g}g^{\alpha\beta}f'(R)\right) + \kappa gg^{\alpha\beta}A_\lambda\right)\delta u^\lambda_{\alpha\beta} + \\
&\quad - \overset{\Gamma}{\nabla}_\mu(gg^{\mu\nu})\delta A_\nu + \overset{\Gamma}{\nabla}_\lambda\left(\frac{\sqrt{g}}{\kappa}g^{\alpha\beta}f'(R)\delta u^\lambda_{\alpha\beta} + gg^{\lambda\nu}\delta A_\nu\right),
\end{aligned} \tag{6}$$

where we used the well-known identity $\delta R_{(\alpha\beta)} = \overset{\Gamma}{\nabla}_\lambda \delta u^\lambda_{\alpha\beta}$ and we set for the energy-momentum tensor $T_{\alpha\beta} := \sqrt{g}\left(g_{\alpha\beta}g^{\mu\nu}\overset{\Gamma}{\nabla}_\mu A_\nu - \overset{\Gamma}{\nabla}_{(\alpha} A_{\beta)}\right)$.

Field equations are

$$\begin{cases}
f'(R)R_{(\alpha\beta)} - \frac{1}{2}f(R)g_{\alpha\beta} = \kappa T_{\alpha\beta}\,, \\[4pt]
\overset{\Gamma}{\nabla}_\lambda\left(\sqrt{g}g^{\alpha\beta}f'(R)\right) = \alpha_\lambda \sqrt{g}g^{\alpha\beta}f'(R)\,, \\[4pt]
\overset{\Gamma}{\nabla}_\mu(gg^{\mu\nu}) = 0\,,
\end{cases} \tag{7}$$

where we set $\alpha_\lambda := -\kappa\frac{\sqrt{g}}{f'(R)}A_\lambda$. Notice that the third equation (that is the matter field equation) is not enough to fix the connection due to the contraction. Notice also that these are more general than field equations of standard $f(R)$ theories due to the rhs of the second equation (that is originated by the coupling between the matter field $A$ and the connection $\Gamma$). Nevertheless one can analyze these field equations along the same lines used in $f(R)$ theories. Let us thence define a metric $h_{\mu\nu} = f'(R)g_{\mu\nu}$ and rewrite the second equation as

$$\overset{\Gamma}{\nabla}_\lambda\left(\sqrt{h}h^{\alpha\beta}\right) = \alpha_\lambda\sqrt{h}h^{\alpha\beta}\,. \tag{8}$$

According to the analysis of EPS-compatibility done in [43] this fixes the connection as

$$\Gamma^\alpha_{\beta\mu} := \{h\}^\alpha_{\beta\mu} - \frac{\kappa}{2f'(R)}\left(h^{\alpha\epsilon}h_{\beta\mu} - 2\delta^\alpha_{(\beta}\delta^\epsilon_{\mu)}\right)a_\epsilon\,, \tag{9}$$

where for notational convenience we introduced the 1-form $a_\epsilon := \sqrt{g}A_\epsilon$. For later convenience let us notice that we have

$$K^\alpha_{\beta\mu} \equiv \Gamma^\alpha_{\beta\mu} - \{h\}^\alpha_{\beta\mu} = -\frac{\kappa}{2f'(R)}\left(h^{\alpha\epsilon}h_{\beta\mu} - 2\delta^\alpha_{(\beta}\delta^\epsilon_{\mu)}\right)a_\epsilon\,. \tag{10}$$

Now we can define the tensor $H^\alpha_{\beta\mu} := \Gamma^\alpha_{\beta\mu} - \{g\}^\alpha_{\beta\mu}$ and obtain

$$H^\alpha_{\beta\mu} = K^\alpha_{\beta\mu} - \frac{1}{2}\left[g^{\alpha\lambda}g_{\beta\mu} - 2\delta^\lambda_{(\beta}\delta^\alpha_{\mu)}\right]\delta_\lambda \ln f'(R) = -\frac{1}{2f'(R)}\left[g^{\alpha\epsilon}g_{\beta\mu} - 2\delta^\epsilon_{(\beta}\delta^\alpha_{\mu)}\right]\left[\kappa a_\epsilon + \delta_\epsilon f'(R)\right], \tag{11}$$

By substituting into the third field equation we obtain

$$\begin{aligned}
&\overset{g}{\nabla}_\mu(gg^{\mu\nu}) + g\left(H^\mu_{\lambda\mu}g^{\lambda\nu} + H^\nu_{\lambda\mu}g^{\mu\lambda} - 2H^\lambda_{\lambda\mu}g^{\mu\nu}\right) = 0, \\
&\Rightarrow H^\nu_{\lambda\mu}h^{\mu\lambda} - H^\lambda_{\lambda\mu}h^{\mu\nu} = 0, \\
&\Rightarrow -\frac{1}{2f'(R)}\left[\left(h^{\nu\epsilon}h_{\lambda\mu} - 2\delta^\epsilon_{(\lambda}\delta^\nu_{\mu)}\right)h^{\mu\epsilon} - \left(h^{\lambda\epsilon}h_{\lambda\mu} - 2\delta^\epsilon_{(\lambda}\delta^\lambda_{\mu)}\right)h^{\mu\nu}\right](\kappa a_\epsilon + \delta_\epsilon f'(R)) = 0, \\
&\Rightarrow -\frac{3}{f'(R)}h^{\nu\epsilon}(\kappa a_\epsilon + \delta_\epsilon f'(R)) = 0, \qquad \Rightarrow a_\epsilon = -\frac{1}{\kappa}\delta_\epsilon f'(R),
\end{aligned} \tag{12}$$

where $\overset{g}{\nabla}_\mu$ is now the covariant derivative with respect to the metric $g$. Hence the matter field $A_\epsilon = \sqrt{g}a_\epsilon = -\frac{\sqrt{g}}{\kappa}\delta_\epsilon f'(R)$ has no dynamics and it is completely determined in terms of the other fields.

We can also express the connection as a function of $g$ alone (or, equivalently, of $h$ alone)

$$\Gamma^\alpha_{\beta\mu} := \{h\}^\alpha_{\beta\mu} + \frac{1}{2}\left(h^{\alpha\epsilon}h_{\beta\mu} - 2\delta^\alpha_{(\beta}\delta^\epsilon_{\mu)}\right)\delta_\epsilon \ln f'(R) \equiv \{g\}^\alpha_{\beta\mu} \tag{13}$$

This behaviour, which has been introduced by the matter coupling, is quite peculiar; the model resembles in the action an $f(R)$ theory but in solution space the connection is directly determined by the original metric rather than by the conformal metric $h$ as in $f(R)$ theories. Still the metric $g$ obeys modified Einstein equations. In fact, we have the first field equation which is now depending on $g$ alone, since the matter and the connection have been determined as functions of $g$. The *master equation* is obtained as usual by tracing (using $g^{\alpha\beta}$)

$$f'(R)R - 2f(R) = \kappa T \qquad \Rightarrow f(R) = \frac{1}{2}(f'(R)R - \kappa T) \tag{14}$$

where we set $T := T_{\alpha\beta}g^{\alpha\beta}$ the trace of the stress-energy tensor. By substituting back into the first field equation, being $T = -\frac{3}{\kappa}\Box f'(R)$, we obtain

$$\begin{aligned}
&f'(R)\left[R_{\alpha\beta} - \frac{1}{4}Rg_{\alpha\beta}\right] - \frac{3}{4}\Box f'(R)g_{\alpha\beta} = \nabla_{(\alpha}\nabla_{\beta)}f'(R) - \Box f'(R)g_{\alpha\beta}, \\
&\Rightarrow R_{\alpha\beta} - \frac{1}{2}Rg_{\alpha\beta} = \frac{1}{f'(R)}\left[\nabla_{(\alpha}\nabla_{\beta)}f'(R) - \frac{1}{4}\left(\Box f'(R) + f'R\right)g_{\alpha\beta}\right],
\end{aligned} \tag{15}$$

where now the curvature and covariant derivatives refer to $g$. These are exactly the field equations obtained in the corresponding purely-metric $f(R)$ theory [24].

Hence we have that, regardless of the function $f(R)$, when there is no matter field other than the field $A$ all these models behave exactly as metric $f(R)$ theories. The conformal factor is $f'(R)$ and $R$ can be calculated in terms of $T$, i.e. $\phi(T) = f'(R(T))$. Field equations then imply that Einstein equations hold for the new metric $\hat{g}$ (corresponding to the above $h$), with a suitably modified stress-energy tensor that takes into account extra effects due to the conformal factor. The previous Einstein equations are recovered

$$\hat{R}_{\alpha\beta} - \frac{1}{2}\hat{R}\hat{g}_{\alpha\beta} = \kappa\hat{T}_{\alpha\beta} \tag{16}$$

with

$$\hat{T}_{\alpha\beta} = \frac{1}{f'}\left[T_{\alpha\beta} + T^{(grav)}_{\alpha\beta}\right], \tag{17}$$

where the first term on the *rhs* is due to a standard matter term- Clearly Eq.(17) means that the extra degrees of freedom coming from $f(R)$ gravity can be managed as a further contribution to the stress-energy tensor and the above observational shortcomings, related to GR (e.g. DM and DE), can be, in principle, solved in a geometrical way.

Which are the physical implications from the EPS formalism point of view?

1. being $g$ and $\hat{g}$ conformally related photon propagation does not change;

2. Einstein equations hold for the new metric $\hat{g}$ with extra stress-energy tensor directly generated by "ordinary" matter $T$;

3. rods and clocks change pointwise, by a factor depending on $T$.

In summary, EPS formalism works also for ETG and further information can be always enclosed in a suitable definition of stress-energy tensor.

## VI. DISCUSSION AND CONCLUSIONS: A NEW PARADIGM FOR GRAVITY

The coupling of a non-linear gravitational Lagrangian $f(R)$ with matter Lagrangians, depending on the metric $g$ in an arbitrary way or even on the connection $\Gamma$ in a peculiar way (dictated by EPS compatibility), generate a set of modified Einstein equations in which the following effects are easily recognizable:

A new metric $\hat{g}$ - conformally related to the original metric $g$ - arises. The conformal factor is a computable function of curvature and, through functional inversion, of the trace of the stress tensor that corresponds to the "ordinary" matter distribution (including possible DM and DE effects). In the Palatini approach, the new metric generates the connection $\Gamma$ as its Levi-Civita connection, so that it describes the free fall of ordinary matter. This new metric induces, in fact, a change of rulers and clocks that affects measurements and conservation laws, while the original $g$ is directly related to light propagation. Due to conformal equivalence, light propagates on the same null geodesics of both $g$ and $\hat{g}$, although clock rates are different in presence of matter.

The net effect of non-linearity and of (non trivial) interaction with matter resides in a change of the stress tensor that couples to the Einstein tensor of $\hat{g}$; a change that induces additions to the previously existing one (directly generated from the matter Lagrangian as discussed above).

This new stress-energy tensor defines conservation laws that are fully covariant with respect to the Einstein frame of $\hat{g}$. Furthermore, it contains an additional term, that can be interpreted under the form of a "space-time varying cosmological constant" $\Lambda(x)$ - in turn determined by distribution of ordinary (and Dark) Matter - so that the residual amount could be interpreted as a net curvature effect (DE) due to the change of rules and clocks induced by EPS compatibility [25]. In other words, the observational effects of such a dynamics are the clustering of astrophysical structures (DM) and the revealed cosmic speed up (DE).

To conclude, we can say that very likely Einstein today, after the new phenomenological evidences would much probably come back onto his own steps and accept, as he always did, that models are not eternal and should be dictated by phenomenology rather than by pre-established rules and prejudices. Why should we insist on pre-judicial rules that impose metricity *a priori* (and metricity with respect to a given metric!) and insist on the choice of the "simplest" Hilbert-Lagrangian, when cosmology, quantum Issues and strings suggest instead to us to strictly follow the beautiful analysis of EPS, and work at least *a priori*, in the extended framework of Palatini-EPS formalism and in a much larger class of Lagrangians?

Moreover, let us remark that working in the extended setting suggested by the Palatini-EPS framework requires to reconsider all the machinery and settings of the observational paradigms and protocols have to be carefully analyzed to disentangle purely metrical effects from effects that measure the interaction with free-fall (and therefore with the connection) that in purely metric formalism GR are necessarily mixed up and entangled by the *a priori* requirement that free-fall is also driven by the metric.